\documentclass{sf2a-conf2025}
\usepackage{graphicx}
\usepackage{hyperref}
\usepackage[]{natbib}  
\usepackage{epstopdf}
\usepackage{xcolor} 

\def\BibTeX{{\rm B\kern-.05em{\sc i\kern-.025em b}\kern-.08em
    T\kern-.1667em\lower.7ex\hbox{E}\kern-.125emX}}
\bibpunct{(}{)}{;}{a}{}{,}  

\definecolor{electricviolet}{rgb}{0.56, 0.0, 1.0} 

\newif\ifcomments   
\commentstrue 
\ifcomments 
  \newcommand{\crmo}[1]{\textcolor{electricviolet}{\textbf{[#1]}}}  
\else   
  \newcommand{\crmo}[1]{}   
\fi 
\newif\ifcomments   
\commentstrue 
\ifcomments 
  \newcommand{\cnl}[1]{\textcolor{green}{\textbf{[#1]}}}    
\else   
  \newcommand{\cnl}[1]{}    
\fi 

\usepackage{eurosym} 
\usepackage{enumitem}

\begin{document}

\TitreGlobal{SF2A 2025}

\title{COMMISSION FEMMES ET ASTRONOMIE DE LA SF2A :\\
WOMEN PARTICIPATION IN FRENCH ASTRONOMY 2025}

\runningtitle{WOMEN PARTICIPATION IN FRENCH ASTRONOMY 2025}

\author{N. Lagarde}\address{Laboratoire d'Astrophysique de Bordeaux, Universit\'e Bordeaux, CNRS, B18N, All\'ee Geoffroy Saint-Hilaire, 33615 Pessac, France}

\author{R.-M. Ouazzani}\address{LIRA, Observatoire de Paris, Universit\'e PSL, Sorbonne Universit\'e, Universit\'e Paris Cit\'e, CY Cergy Paris Universit\'e, CNRS, 92190 Meudon, France}

\author{J. Malzac}\address{IRAP, Universit\'e de Toulouse, CNRS, UPS, CNES, Toulouse, France.}

\author{M. Clavel}\address{Univ. Grenoble Alpes, CNRS, IPAG, F-38000 Grenoble, France}

\author{P. de Laverny}\address{Universit\'e C\^ote d'Azur, Observatoire de la C\^ote d'Azur, CNRS, Laboratoire Lagrange, Bd de l'Observatoire, CS 34229, 06304 Nice
cedex 4, France}

\author{L. Leboulleux$^4$}

\author{I. Vauglin}\address{CRAL-Observatoire de Lyon, CNRS, UCBL, ENS-Lyon, 69561 Saint-Genis Laval Cedex, France}

\author{C. Bot}\address{Universit\'e de Strasbourg, Observatoire Astronomiques de Strasbourg}

\author{S. Brau-Nogu\'e$^3$}

\author{L. Ciesla}\address{Aix Marseille Univ, CNRS, CNES, LAM, Marseille, France}

\author{E. Josselin}\address{LUPM, Universit\'e de Montpellier, CNRS, Place E.Bataillon, 34095 Montpellier, France}

\author{N. Nesvadba$^5$}

\author{O. Venot}\address{Universit\'e Paris Cit\'e and Univ Paris Est Creteil, CNRS, LISA, F-75013 Paris, France}

\setcounter{page}{237}

\maketitle

\begin{abstract}
\textit{The Commission Femmes et Astronomie} of the French Astronomical Society, has conducted a statistical study aimed at mapping the current presence of women in French professional astronomy and establishing a baseline for tracking its evolution over time. This study follows an initial survey carried out in 2021, which covered eight astronomy and astrophysics institutes (1,060 employees). This year, the scope was expanded to 11 institutes, bringing together a total of 1,525 employees, including PhD students, postdoctoral researchers, academics, as well as technical and administrative staff, representing about 57\% of the whole French community.
We examined how the proportion of women varies according to career stage, level of responsibility, job security, and income. The results are compared to the 2021-2022 survey and appear to illustrate the well-known "leaky pipeline", with one of the main bottlenecks being access to permanent positions. The study shows that the proportion of women consistently declines with increasing job security, career seniority, qualification level, and salary.

\end{abstract}


\section{Introduction}
The \textit{Commission Femmes et Astronomie} of the \textit{SF2A} (\textit{Soci\'et\'e Fran\c caise d'Astronomie et d'Astrophysique}, French astronomical Society) was created in 2020 to form an instance where questions related to gender equality can be addressed within the French astronomical community. The Commission has twelve members, of which six members are currently --or were at some point-- also part of the SF2A Council: Caroline Bot, Sylvie Brau-Nogu\'e, Laure Ciesla, Patrick de Laverny, Eric Josselin, Nad\`{e}ge Lagarde*, Lucie Leboulleux, Rhita-Maria Ouazzani*, Nicole Nesvadba, Julien Malzac*, Isabelle Vauglin, and Olivia Venot\footnote[1]{Current members of the SF2A Council}. The main goals of the Commission are to promote gender equality in Astronomy \& Astrophysics in France, fight against sexual and gender-based violence and support gender-focused outreach actions.

Before the commission's creation, several efforts had already been made to document the status of women in French astronomy. A survey of young researchers (PhD graduates in Astronomy and Astrophysics between 2007 and 2017) revealed that women were less likely than men to obtain permanent positions \citep{Berne2020}. In parallel, a census of permanent staff found that 23\% of permanent positions were held by women, with contrasting trends depending on the type of position: a decrease for university positions but an increase for astronomer (CNAP) ones, with no evidence of a glass ceiling \citep{BotBuat2020}. These studies were extremely relevent and well conducted, but i) they were biased by the surveyed population, and i) they gave an instantaneous glimpse of the status of women in astronomy in 2019-2020.

In this context, the \textit{Commission Femmes et Astronomie} conducted a statistical study that aimed at mapping the presence of women in French professional Astronomy in 2021, based on standard information provided by a subset of INSU-Astronomy\footnote{The Astronomy and Astrophysics perimeter of the {\it Institut National des Sciences de l'Univers}, CNRS} Institutes directly. The results of the study illustrated the leaky pipeline, with one major bottleneck being the access to permanent positions. In general, the 2021 study showed that the proportion of women steadily decreases with job stability, with the career stage, with the qualification level and with the income level.
The aim of the present work is to take this as the starting point of a series of surveys, conducted regularly, allowing to study the evolution of inclusion of women in French Astronomy over time. 

With this series of studies, we seek to address the following questions:
\begin{itemize}
    \item How has the proportion of women in our institutes, observatories, and research areas evolved over time?
    \item How has the distribution of women across career levels, positions of responsibility, and income changed with time?
    \item How have women's career trajectories developed and transformed over the years?
    \item What barriers continue to hinder their progress?
    \item How these obstacles evolve in time?
\end{itemize}

The ultimate goal is to draw a diagnostic assessment of gender inequalities within our community, grounded in quantitative data, allowing to identify structural or conjuncture's factors, and hence suggest efficient policies to improve diversity and inclusion in our field.

\section{The 2024-2024-2025 Map}

\subsection{The 2024-2025 study set up}

We collected data from eleven research institutes of the INSU-AA$^{*}$ : IRAP, LESIA, GEPI, LUTH, LAGRANGE, IAS, IPAG, LAB, LAM, CRAL and UTINAM, comprising 1,525 individuals, including scientific, technical and administrative staff, with all of the possible employers, which represents about 57\% of the community. This work represents an increase in the number of institutes surveyed compared to our previous study \citep{Ouazzani22}, which included 8 institutes, and thus an increase in the number of staff members covered. For this survey, we have 5 institutes in common with the previous one, namely GEPI, IRAP, LAGRANGE, LESIA, and LUTH. For this common set of institutes, the sample size was 764 in 2021-2022 compared with 844 in 2024-2025. With the sample being roughly comparable, or at least slightly larger, we can already observe a decrease in both the proportion and the number of women between the two studies. Specifically, the proportion of women fell by 4\%, (- 6 in absolute number) while the number of men clearly increased (+ 87). 

For all these institutes surveyed in 2024-2025, the data set contained the following entries\footnote{Data marked with an asterisk in the original data set were collected with varying degrees of completeness.}:

\begin{itemize}[label=-,noitemsep, topsep=0pt]
    \item institute
    \item Gender
    \item Age group
    \item Employer (CNRS, CNAP, University, OSU)
    \item Status (PhD student, ITA\footnote{ITA stands for \textit{Ing\'enieur$\cdot$es, Technicien$\cdot$nes, Administratif/ves}, which translates to: engineering, technical and administrative staff}, researcher/lecturer, postdoc)
    \item Year of joining the unit/institute
    \item Corps (civil service category)
    \item Grade$^*$
    \item BAP (professional ITA's branch)$^*$
    \item Seniority in grade$^*$
    \item HDR (Accreditation to Supervise Research)$^*$
    \item HDR date$^*$
    \item Work quota (employment percentage)$^*$
\end{itemize}

\subsection{Results for the general population with a focus on job security}

\begin{figure}[t!]
 \centering
 \includegraphics[width=0.45\textwidth,clip]{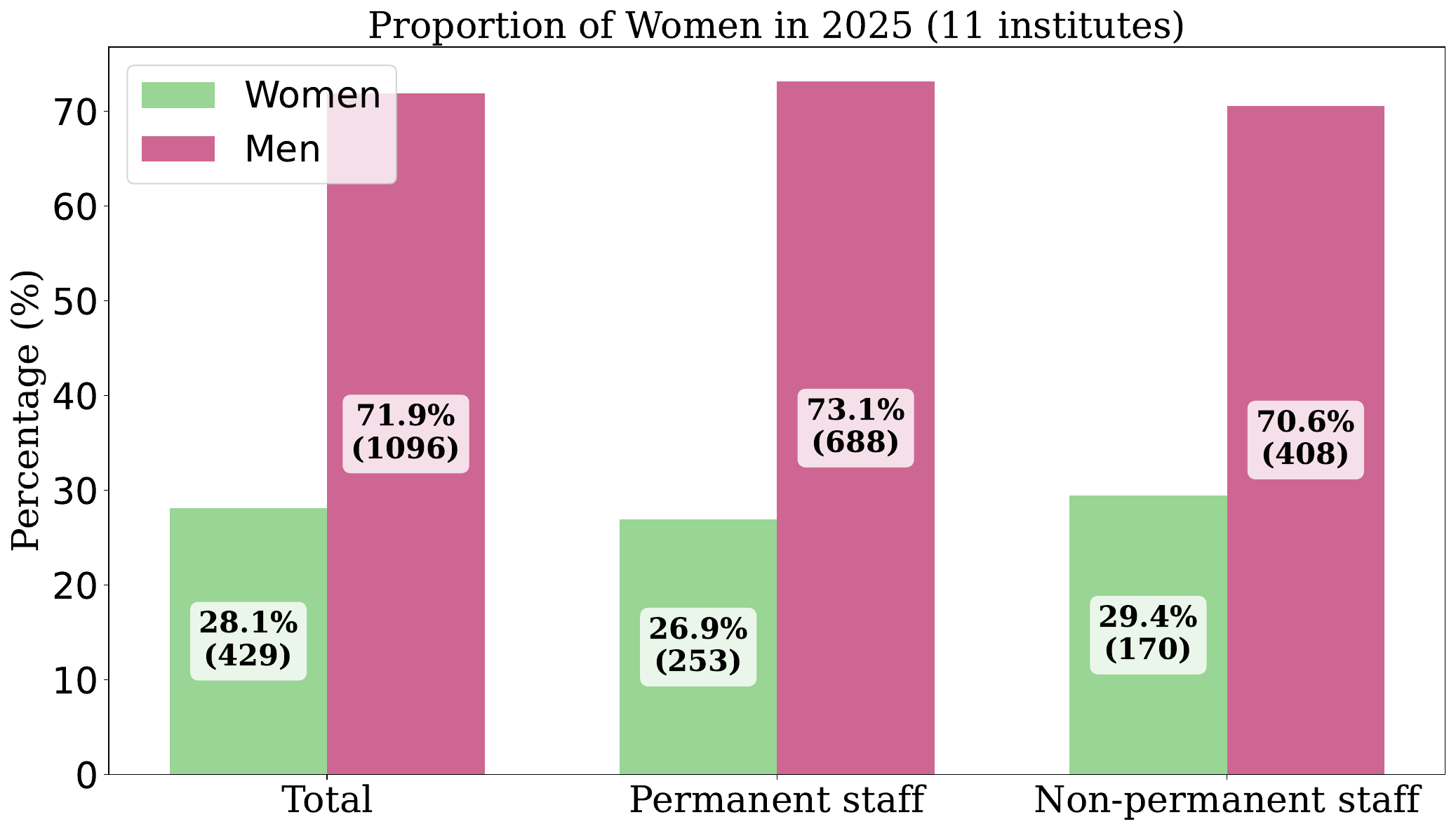}
 \includegraphics[width=0.45\textwidth,clip]{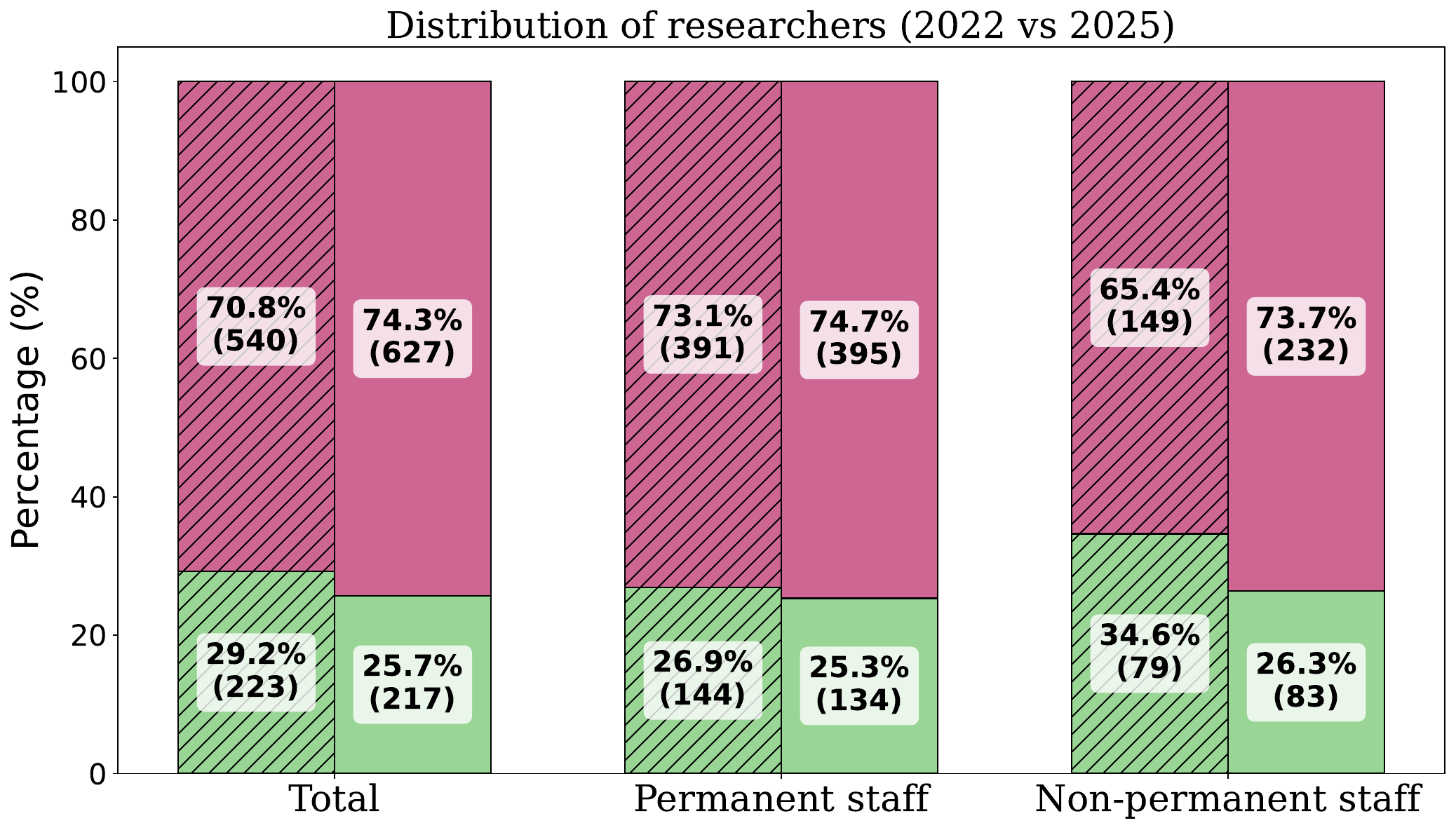}    
  \caption{Proportion of women for the overall sample, among permanent staff, and non-permanent staff. Left:  based on the complete 2024-2025 sample. Right:  restricted to the five laboratories included in both the 2021-2022 (hatched bars) and 2024-2025 studies.}
  \label{lagarde:fig1}
\end{figure}

The proportion of women in the total sample collected in 2024-2025 is shown in Figure~\ref{lagarde:fig1} , according to their employment status and position. The workforce in French academia is composed primarily of permanent staff, including for a vast majority civil servants in research, engineering, technical, or administrative roles, as well as a small but growing number of employees on private sector like permanent contracts. Non-permanent positions include PhD students, postdoctoral researchers, teaching assistants, apprentices, and holders of short-term contracts in research, engineering, technical, or administrative functions. 

Examining the distribution of women across permanent and non-permanent positions (Fig.\ref{lagarde:fig1}, left), the proportion of women in non-permanent position is higher (29.4\%) than that of women in permanent positions (26.9\%), while it's the opposite for men (permanent 73.1\% versus non-permanent 70.6\%). In other words, we observe that women (39.6\%) are more likely than men (37.2\%) to hold temporary contracts. Considering only the institutes in common with the 2021-2022 study (Fig.\ref{lagarde:fig1}, right), we notice a slight decrease in the proportion of women in permanent positions (from 26.9\% to 25.3\%) and a clear increase in the proportion of men in temporary positions (from 65.4\% to 73.7\%).\\

\begin{figure}
 \centering
 \includegraphics[width=0.45\textwidth,clip]{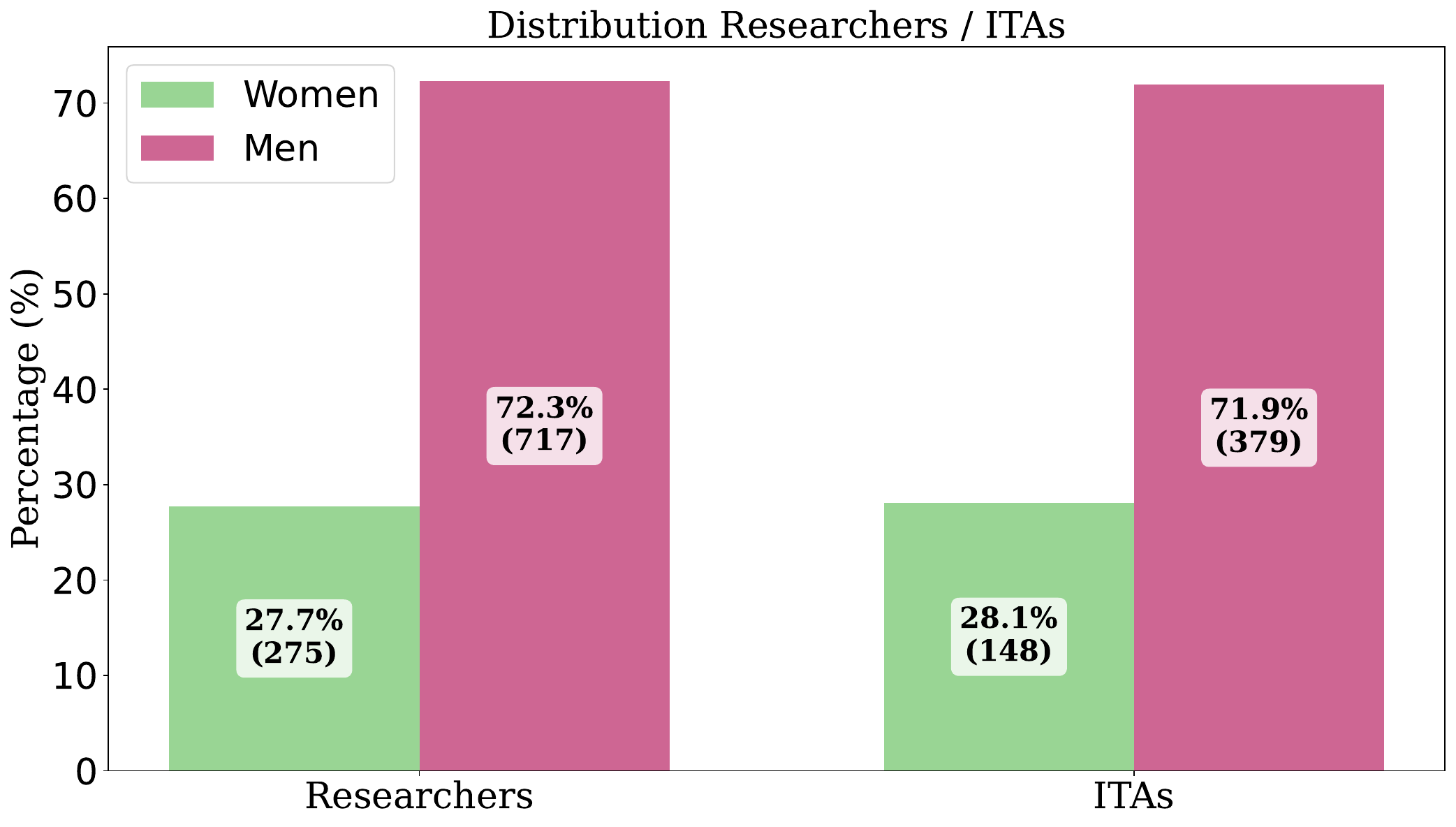}
 \includegraphics[width=0.45\textwidth,clip]{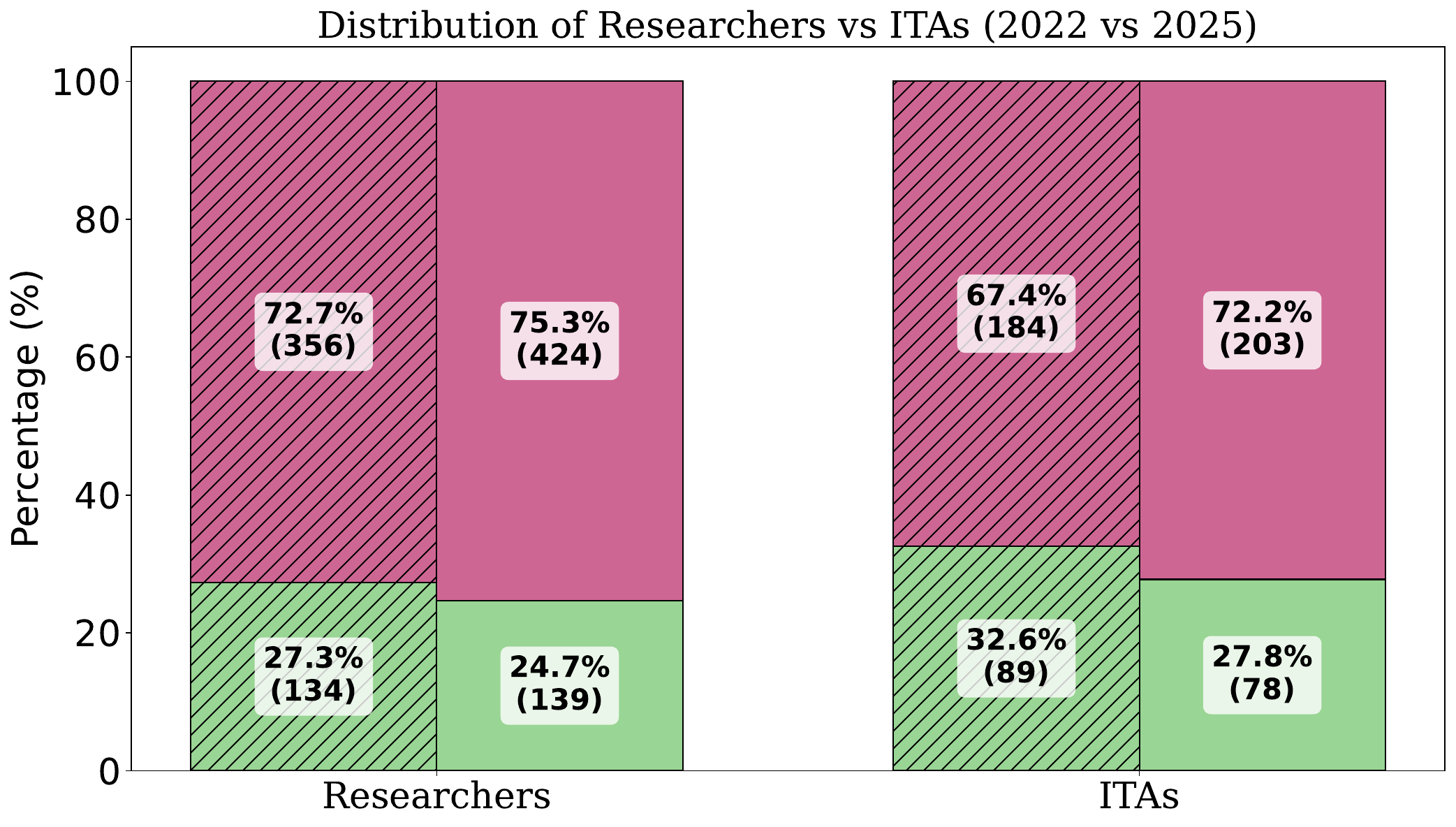}    
  \caption{Proportion of women among the researchers (from PhD to Emeritus, 992 individuals), and among administrative, technical and engineering staff (a.k.a ITA, 527  individuals). Left:  based on the complete 2024-2025 sample. Right:  restricted to the five laboratories included in both the 2021-2021-2022 (hatched bars) and 2024-2025 studies. }
  \label{lagarde:fig2}
\end{figure}

One potential source of variability in these numbers is expected to arise from the type of position (research or non-research), which is explored in the Fig. \ref{lagarde:fig2}. We acknowledge that individuals in ITA positions also contribute to the scientific outcome of these institutes; however, research and ITA functions are ascribed distinct symbolic, social, and economic values, which we believe is a key factor here: the proportion of women ITA is slighly larger (28.1\%) than that of women in research positions (27.7\%). Among women employed in the 11 institutes included in this study, 35.0\% hold ITA positions, compared with 34.6\% for men. These proportions were similar in the 2021-2022 study (see Fig. \ref{lagarde:fig2}, right). For the subsequent analyses, we treat separately the population of researchers (in the broad sense, from PhD students to Emeritus researchers) and the population of Engineering, Technical, and Administrative staff (ITA).

\subsection{Research career}

\begin{figure}[t!]
 \centering
 \includegraphics[width=0.45\textwidth,clip]{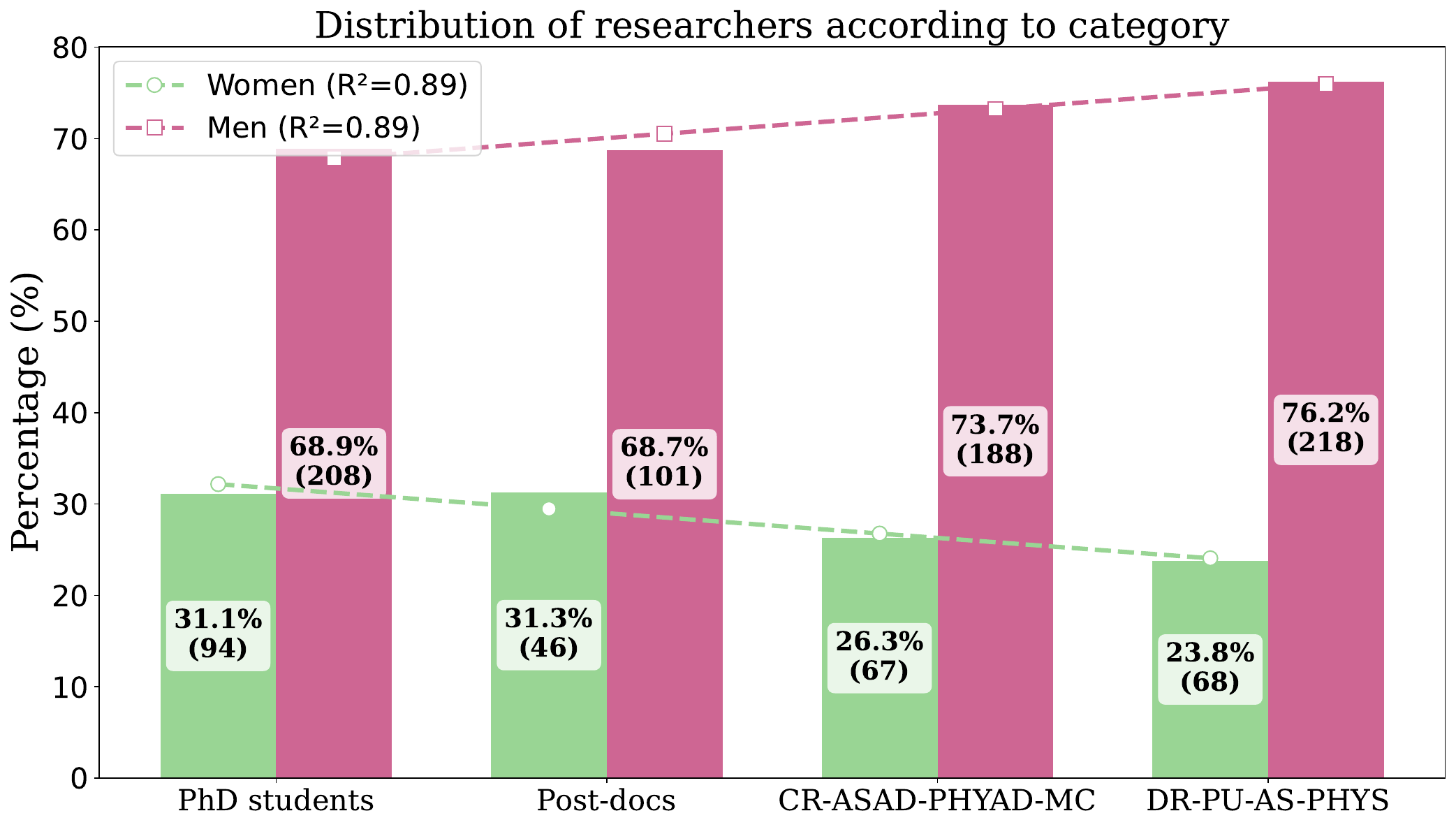}  
 \includegraphics[width=0.45\textwidth]{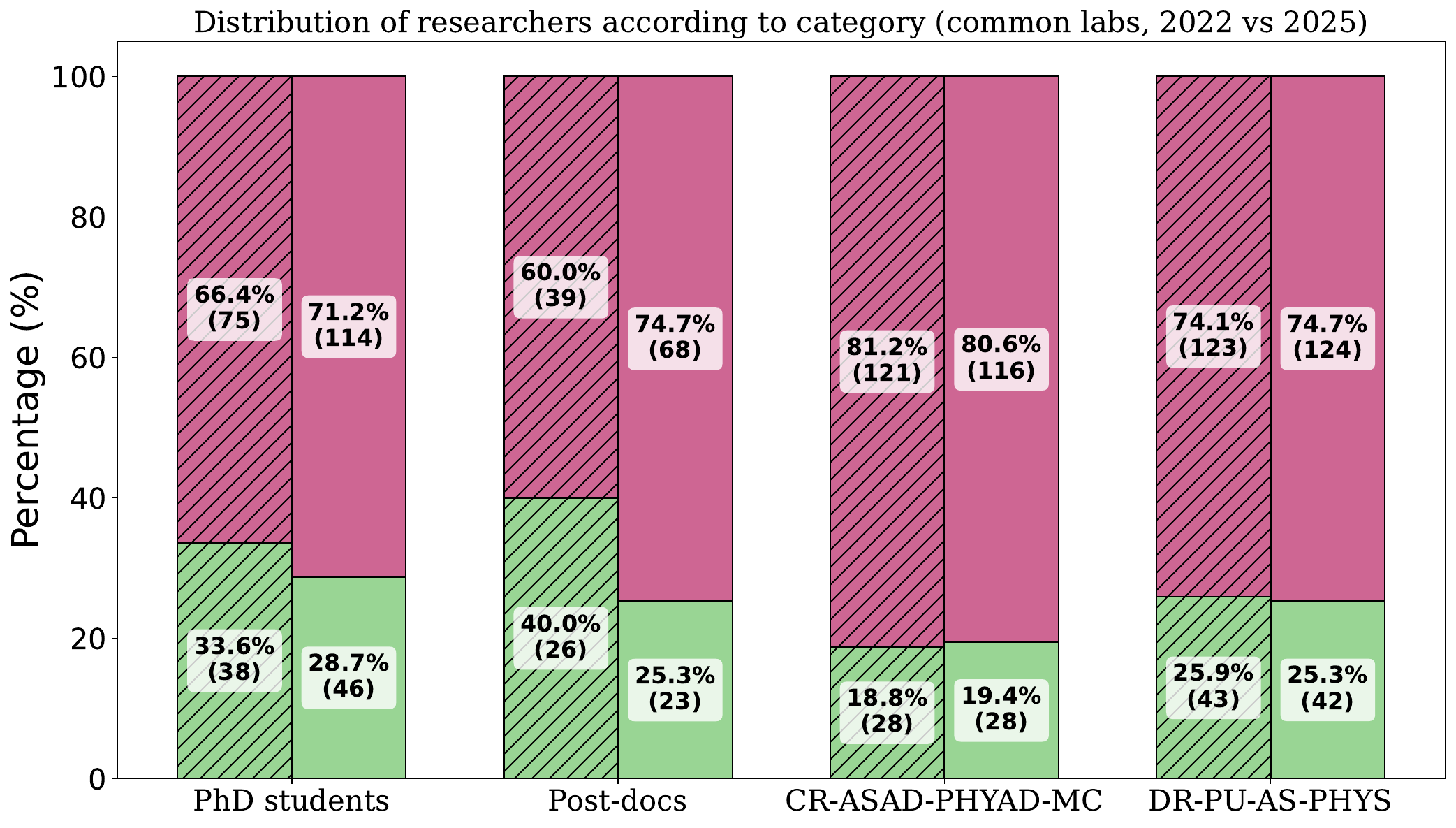}  
  \caption{Proportion of women at each stage of the research career. Left panel: based on the complete 2024-2025 sample, with dashed lines showing linear fits for women (green) and men (magenta). Right panel: restricted to the five laboratories included in both the 2021-2022 (hatched bars) and 2024-2025 studies. }
  \label{Lagarde:fig3}
\end{figure}

We now focus on research positions only, at all career stages, starting from PhD students to emeritus researchers (992 individuals). Concerning job security, we find that women are generally more likely than men to hold temporary positions: 50.9\% of women hold temporary contracts versus 40.4\% of men (45.5\% for the whole researchers population).
Concerning career stages, by increasing seniority, we considered: PhD students, postdoctoral researchers, positions equivalent to associate professors (CR, ASAD, PHYSAD, MC), second-class professors (DR2, PU2, AST2), first-class professors (DR1, PU1, AST1), professors of exceptional class (DRCE, PUCE, ASTCE), and Emeritus researchers\footnote{for readers who are not fluent in French administrative jargon, we refer to the appendix of \cite{Ouazzani22}}. The distribution is presented in Fig.\ref{Lagarde:fig3}. It clearly illustrates a pronounced decline in the number of women as seniority increases. To highlight the overall trend, a linear fit was applied, revealing that the proportion of women decreases with career stage, typical of what we call the leaky pipeline. As was noted in the first study, the leakage is particularly striking at the transition from non-permanent to permanent positions: women represent about $\sim$31\% of PhD students and postdoctoral researchers ($\sim$35\% in 2021-2022), but only 26.3\% at the first level of permanent employment (23.4\% in 2021-2022), decreasing further to 23.8\% at the professor/astronomer/research director level. Note that if we consider the highest rank positions in the associate professor level (\textit{hors classe:} CRHC, MCHC, ASADHC), the proportion of women is very low, at 17.6\%.

If we want to explore how these numbers evolved since 2021-2022, we restrict ourselves to the institutes included in both studies (2021-2022 and 2024-2025). As illustrated in the left panel of Fig.\ref{Lagarde:fig3}, the number of permanent researchers has remained particularly stable (71 women and 244 men in 2024-2025, against 70 women and 240 men in 2021-2022), with proportions unchanged. However, there is a substantial decrease in the share of women in short-term positions, driven by a marked increase in male recruitment in those positions. In fact, the number of women in non-permanent positions slightly increased (from 64 in 2021-2022 to 69 in 2024-2025), while, simultaneously, the number of male increased drastically from 114 to 183. This net increase in the number of PhD (+39) and postdoctoral (+29) positions show that recruitment between 2021-2022 and 2024-2025 in short-term contracts has largely favored male early career researchers.

As in many other countries, career progression in France toward the highest \textit{corps}; namely full professor, research director, or full astronomer; is not automatic but competitive. To be eligible, candidates are generally expected to hold an Accreditation to Supervise Research (\textit{Habilitation \`a Diriger des Recherches}, HDR). While not formally mandatory, in practice virtually all successful applicants obtain the HDR beforehand. Figure \ref{Lagarde:fig3} (left) suggests a second bottleneck after the transition from postdoctoral to permanent positions: women account for 26.3\% of associate professor-level positions, but only 23.8\% at the full professor level. This raises the question of whether the leakage between these stages is linked to fewer women holding the HDR at the lower level. Although data were incomplete (only 193 individuals across 11 institutes), the figures are telling: among associate professor-level staff, 39.4\% (86) hold the HDR. Disaggregated by gender, 25.5\% of women (12/47) are accredited, compared to 43.8\% of men (64/146). Overall, women represent just 18.8\% of those accredited. This is troubling, given the importance of the HDR for promotion.

To fully establish whether the underrepresentation of women at the full professor level is caused by a shortage of HDR holders among women, we would also need to exclude the possibility that accredited women are promoted more rapidly and thus spend less time at the associate professor level. Testing this hypothesis would require correlating the date of HDR completion with promotion dates, however this information not available in the present dataset.

\subsection{Women in engineering and technical positions}
\begin{figure}[t!]
 \centering
  \includegraphics[width=0.45\textwidth,clip]{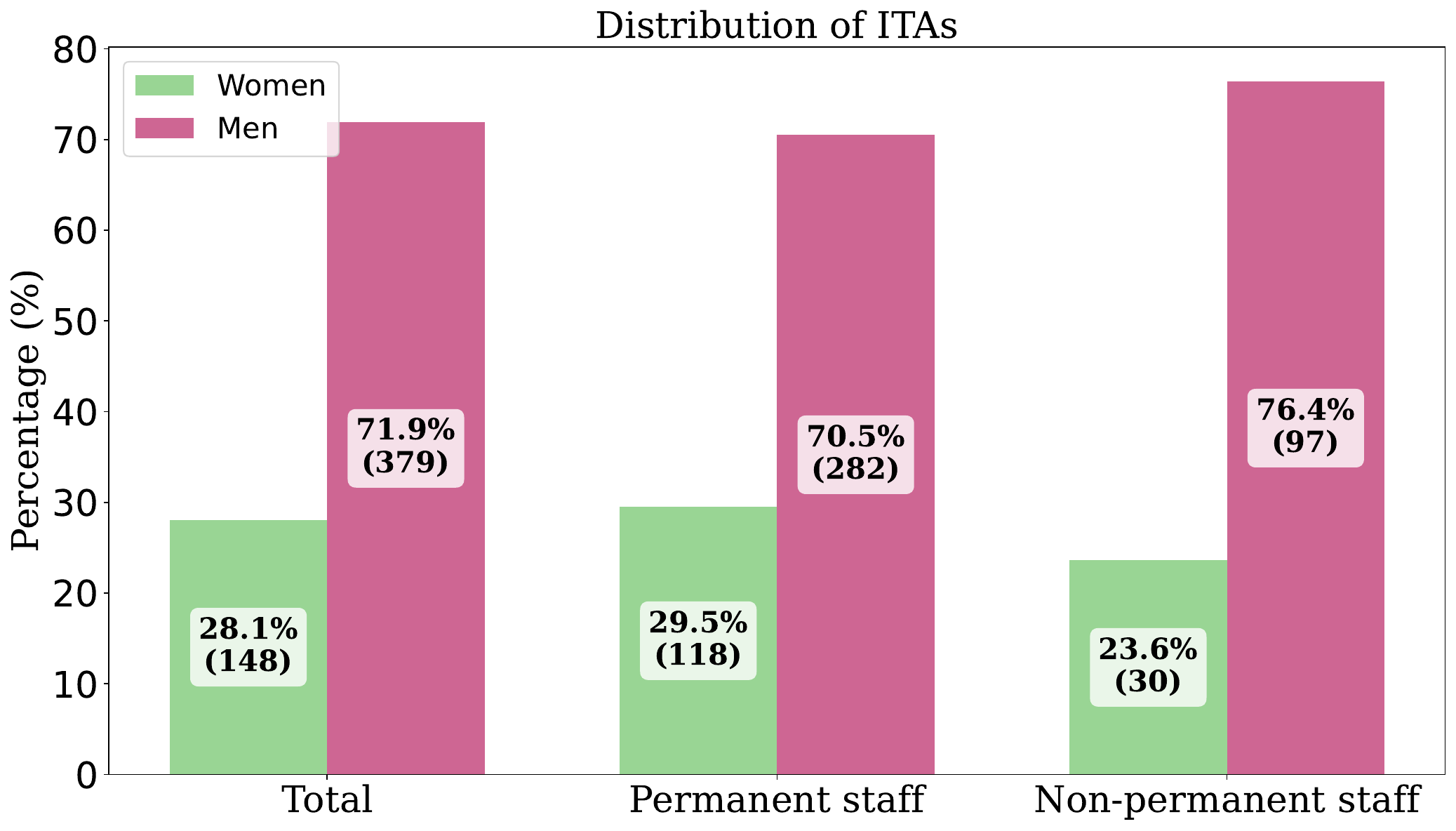}
  \includegraphics[width=0.45\textwidth,clip=true,trim= 0cm 0cm 0cm 0cm]{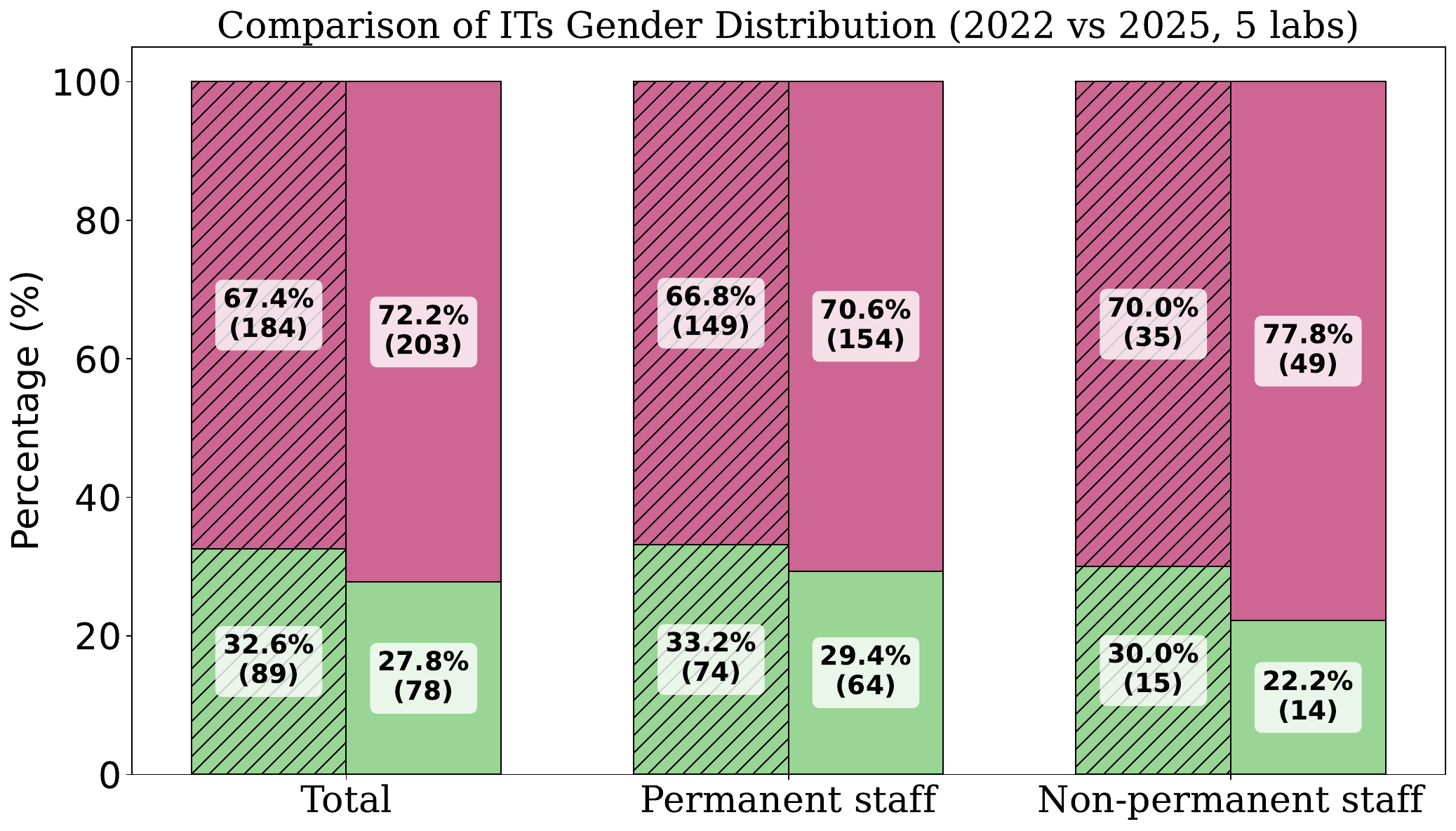}
  \caption{Proportion of women in the ITA population among permanent and non-permanent staff, based on the full 2024-2025 sample (left panel), and restricted to the five laboratories common to both the 2021-2022 (hatched bars) and 2024-2025 studies (right panel). }
  \label{Lagarde:fig4}
\end{figure}

Concerning the ITA population, which includes 527 individuals, the fraction of women holding non-permanent positions is 23.6\%, compared with 29.5\% in permanent positions (see Fig.~\ref{Lagarde:fig4}, left). Again, as for the population of researchers, the proportion of women increases with precariousness. Considering the two studies, with the sample reduced to the 5 institutes in common, there is a 13\% decrease in the share of women ITA in permanent positions, while the number of women ITA with short-term contracts has remained stable (see Fig.~\ref{Lagarde:fig4}, right). In contrast, the fraction of men ITA temporary positions increased markedly, from 70\% in 2021-2022 to 77.8\% in 2024-2025.

\begin{figure}[ht!]
 \centering
 \includegraphics[width=0.45\textwidth,clip]{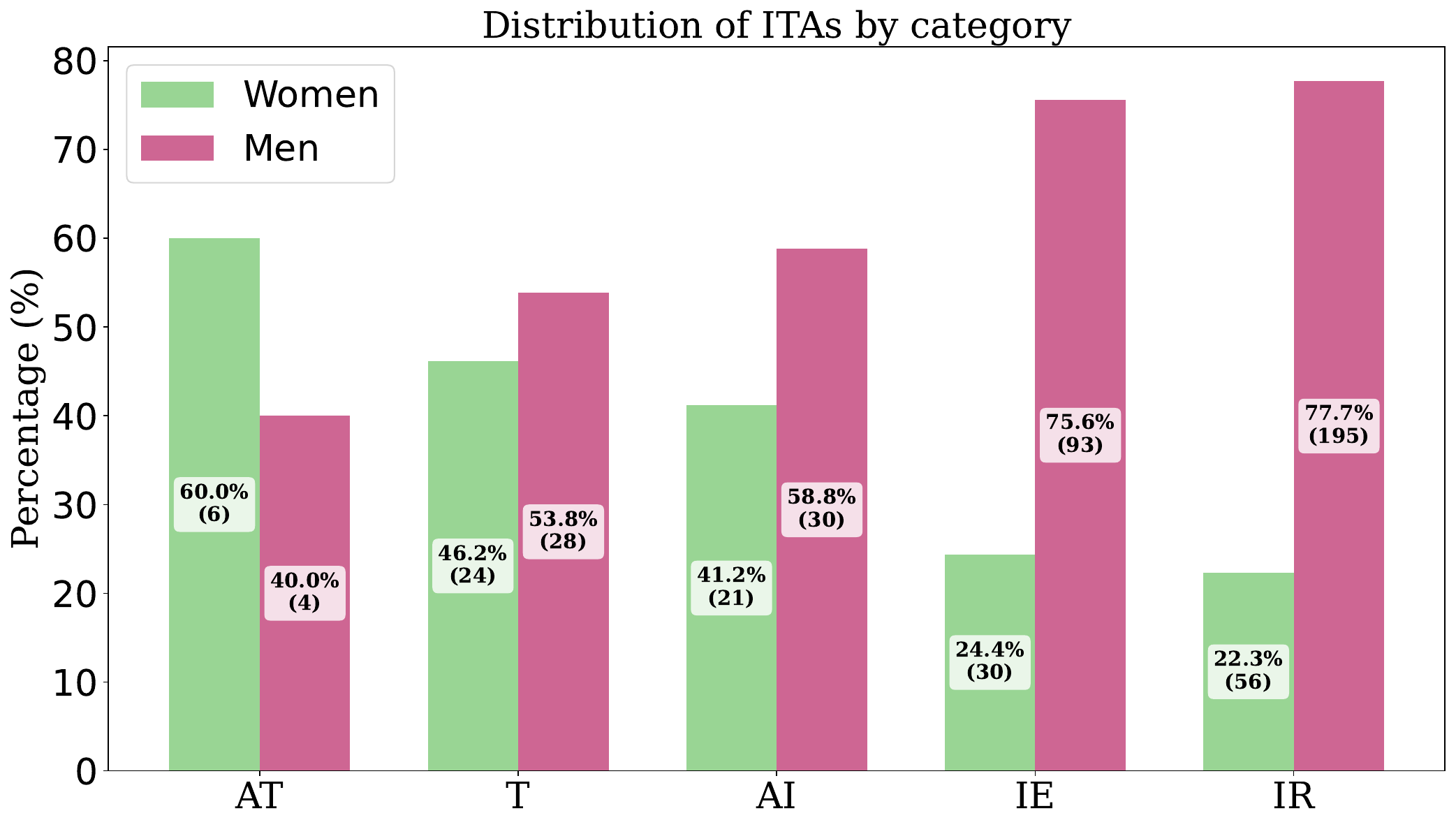} 
 \includegraphics[width=0.45\textwidth,clip]{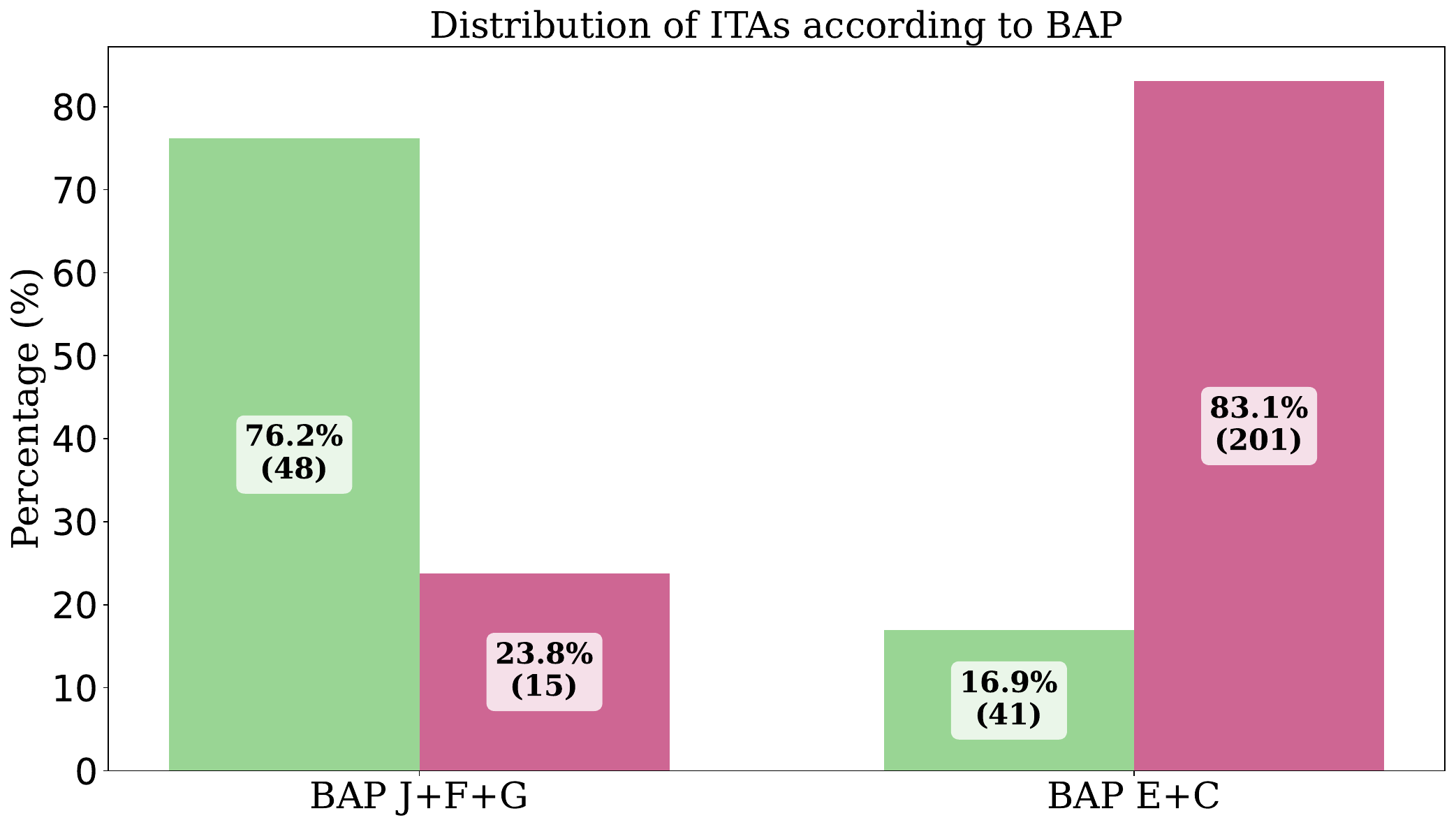}
  \caption{ Left : Proportion of women in each category of ITA jobs. From left to right they are ordered by qualification and income level. Right : Proportion of women in two groups of professional domains : Management, Culture/Communication, Logistics (BAP J, F, G) and IT, Engineering and Scientific Instrumentation (BAP E, C)}
  \label{Lagarde:fig5}
\end{figure}

The distribution of ITA individuals in the full 2024-2025 sample by career level is shown in the left panel of Fig. \ref{Lagarde:fig5}. It should be noted that the histogram groups ITA positions into broad categories (known as \textit{corps} in French), ordered by increasing levels of income and qualification. Unlike for researchers, the progression from one category to another is minimal or nonexistent. Hence, in Fig.~\ref{Lagarde:fig4} the individuals do not navigate from one \textit{corps} to another like in Fig.~\ref{Lagarde:fig3}, and does not illustrate a leaky pipeline. However,  Fig.~\ref{Lagarde:fig5} shows, again, that the more the position is prestigious, socially and financially rewarded, the less women there are.

In this case, there is another source of variability compared to research careers: the nature of the work, or expertise, which can be administrative, technical, or scientific (R\&D, laboratory work or data science to cite a few). Positions in the AT, T, and AI categories are predominantly administrative, whereas IE and IR positions are mainly scientific and technical. Another factor of variability is the level of qualifications required to hold such positions: some require a PhD (IR), while others require a High School Leaving Certificate (French \textit{Baccalaur\'eat}). In summary, Fig. \ref{Lagarde:fig5} (left panel) illustrates that as qualification and income levels increase, the proportion of women decreases steadily, from 60\% in AT positions to only 22.3\% in IR positions.

ITA staff are assigned to different professional branches (BAPs\footnote{in french : Branches d'Activit\'es Professionnelles}) according to their area of expertise. These branches gather positions with similar functions and skill requirements. Specifically: 
\begin{itemize} 
\item BAP J covers management and administrative coordination functions; 
\item BAP F includes positions related to culture, communication, knowledge production, and dissemination; 
\item BAP G encompasses tasks in real estate management, logistics, catering, and safety/prevention; 
\item BAP E includes positions in information technology, statistics, and scientific computing; 
\item BAP C corresponds to engineering and scientific instrumentation.
\end{itemize}

Assigning ITA to BAPs allows the French administration to categorize positions based on both functional responsibilities and required expertise. According to this classification, the right panel of Fig. \ref{Lagarde:fig5} shows the proportion of women across two groups of BAPs. The first group combines BAP J, F, and G, which correspond primarily to management, culture/communication, and logistics positions, while the second group includes BAP E and C, representing engineering and scientific instrumentation. The distribution reveals a marked gender imbalance between these two groups: women account for 76.2\% of positions in the first group, indicating a strong predominance in administrative, cultural, and logistical roles, whereas they represent only 16.9\% of positions in the second group, which are more technical and scientific in nature. These statistics clearly illustrate that, within the French community, some positions remain highly gendered. This is particularly significant given that women are less present in ITA positions as the work becomes more closely connected to science, while men continue to dominate the symbolic hierarchy of the scientific community.

\subsection{The distribution of income across jobs and gender}

\begin{figure}[ht!]
 \centering
 \includegraphics[width=0.8\textwidth]{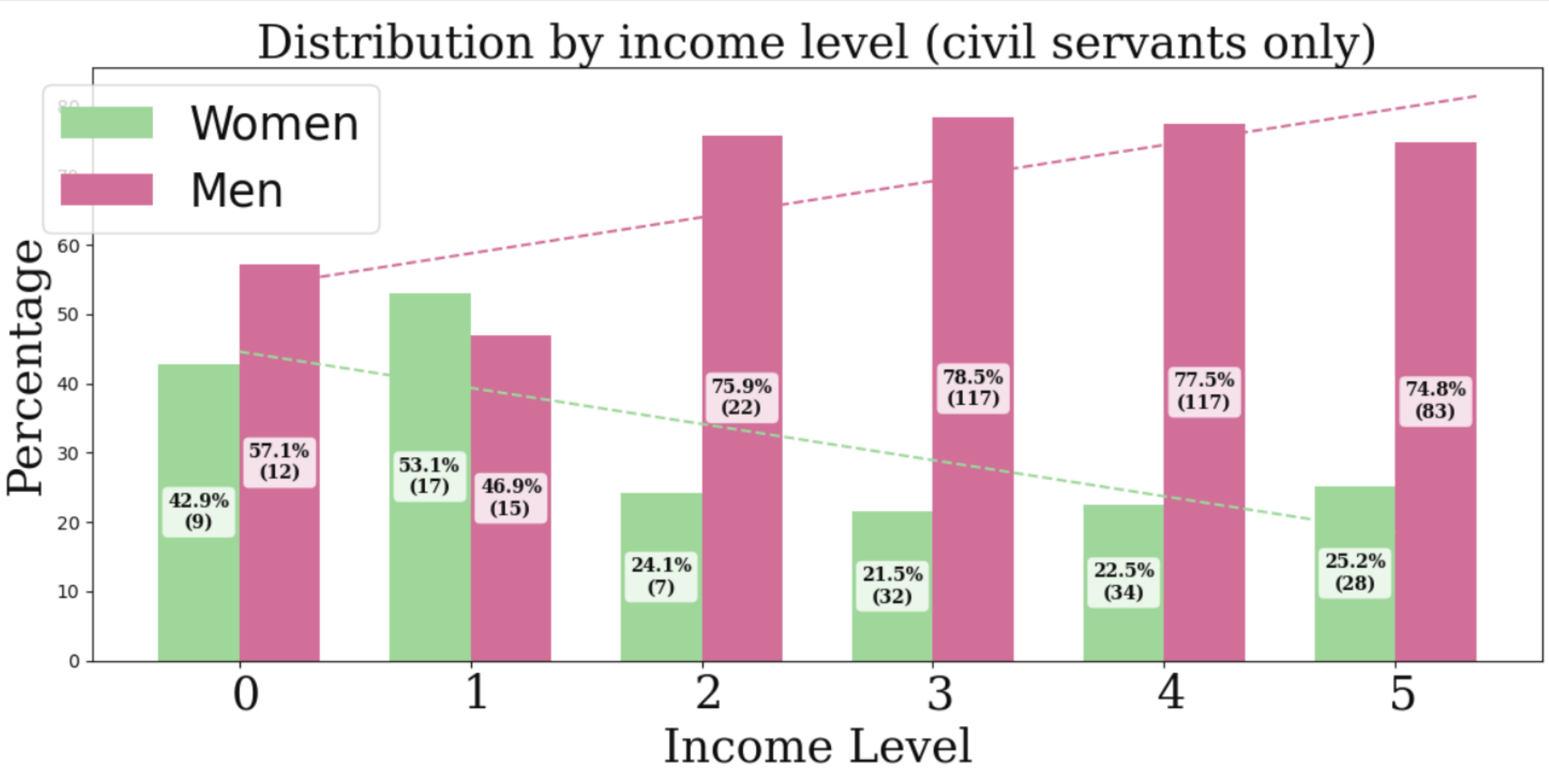}
  \caption{Proportion of women and men at each income level given in Table \ref{table1}, increasing from left to right.}
  \label{Lagarde:fig6}
\end{figure}

In this section, we want to address a key aspect: the distribution of income with respect to gender. For this analysis, we considered the entire data set of 1,525 individuals and classified them according to income levels, as summarized in Table \ref{table1}. While salary is not necessarily a primary motivation for choosing an academic career, it serves as a useful proxy for the social value and responsibility associated with a given position. This is illustrated in Fig.~\ref{Lagarde:fig6}. The lowest two income levels primarily include technical and administrative positions, which are known to display pronounced gender imbalances. For these two first levels, men and women are almost equally distributed. Then, beyond that point, we observe a drop of the share of women which amount to around 25\% for all of the remaining higher levels of income, highlighting the under representation of women in the higher-paid academic positions.

\begin{table*}
\centering
\caption{Scale of income for all the workers in the sample, researchers and ITA altogether. Note: Reported salaries do not include bonuses awarded during the career}
 \label{table1}
 \begin{tabular}{clc}
  \hline
  \hline
 Level & Category & Gross monthly salary \\
0 & AJT2, AJT1, TCN, TCS & $<$2300~\euro\\
1 & AI, TCE & $<$2500~\euro\\
2 & IECN, IR2 & $<$3000~\euro\\
3 & ASADCN, CRCN, MCCN, IEHC, IR1 & $<$4000~\euro\\
4 & ASC2, DR2, PU2, ASADHC, MCHC, IRHC, CRHC &  $<$6000~\euro\\
5 & ASC1, DR1, PU1, ASCE, PUCE, DRCE, Emeritus &  $>$6000~\euro\\
 \hline
 \end{tabular}
\end{table*}

\section{Conclusions}

We present the findings of our second statistical survey of the participation of women in French astronomical institutes, building on an initial study conducted in 2021-2022 by \cite{Ouazzani22}.  In contrast to the first edition, the 2024-2025 survey includes additional information—such as the professional branch for ITA staff and the Accreditation to Supervise Research (HDR), which enables us to address new questions. The main strength of this second edition, however, lies in its ability to track the evolution of several key indicators over time. Our 2024-2025 sample comprises 1,525 individuals from 11 institutes, representing around 57\% of the targeted population.

They indicate that the proportion of women steadily decreases with job security, career stage, qualification level, and income. The comparative analysis of the two data collection campaigns highlights several notable trends in the gender distribution within the French professional astronomy community. The most striking one is the significant increase in the number of men in temporary positions, both among ITA staff and researchers, mainly due to the recruitment of postdocs and PhD students. In particular, although the proportion of women in permanent positions has remained stable at around 22\% over recent years, with near gender-balanced recruitment in the latest national campaigns (CNAP and CNRS), men still outnumber women overall. This reflects the well-known "leaky pipeline" phenomenon in career progression. It is worth noting that, while awareness of gender balance has improved at the national level, earlier recruitment periods were less attentive to this issue, and local recruitments involving only a single available position often continue to favor men. The imbalance is particularly striking among postdoctoral positions, but can also be observed in permanent roles such as university lecturers. 
Moreover, the distribution of men and women ITA staff across BAP categories remains strongly gendered, and the proportion of women consistently decreases with higher income levels, across all job types. These results emphasize the importance of continuing longitudinal monitoring with complete and homogeneous data across laboratories. Establishing a standardized data extraction procedure will be crucial for future editions of such a survey.

\begin{acknowledgements}
The members of the Commission Femmes et Astronomie would like to thank all institute directors and colleagues who kindly agreed to provide the data that made this study possible, as well as the administrative staff who worked on compiling and anonymasing. 
\end{acknowledgements}

\bibliographystyle{aa}  
\bibliography{Lagarde_S00} 

\end{document}